\documentclass[a4paper,twoside]{article}

\usepackage[utf8]{inputenc}
\usepackage{epsfig}
\usepackage{subcaption}
\usepackage{calc}
\usepackage{amssymb}
\usepackage{amstext}
\usepackage{amsmath}
\usepackage{amsthm}
\usepackage{multicol}
\usepackage{apalike}
\usepackage{url}
\usepackage{placeins}
\usepackage{listings}

\usepackage{SCITEPRESS}     

\usepackage
[
a4paper,
left=2.6cm,
right=2.6cm,
top=3.3cm,
bottom=4.2cm
]
{geometry}

\begin{document}

\title{Astronomical Images Quality Assessment with Automated Machine Learning}

\author{\authorname{Olivier Parisot, Pierrick Bruneau, Patrik Hitzelberger}
\affiliation{Luxembourg Institute of Science and Technology (LIST) \\ 5 Avenue des Hauts-Fourneaux, 4362 Esch-sur-Alzette - Luxembourg}
\email{olivier.parisot@list.lu}
}



\abstract{  
Electronically Assisted Astronomy consists in capturing deep sky images with a digital camera coupled to a telescope to display views of celestial objects that would have been invisible through direct observation. This practice generates a large quantity of data, which may then be enhanced with dedicated image editing software after observation sessions.
In this study, we show how Image Quality Assessment can be useful for automatically rating astronomical images, and we also develop a dedicated model by using Automated Machine Learning.
}

\onecolumn \maketitle \normalsize \setcounter{footnote}{0} \vfill

\section{\uppercase{Introduction}}
\label{sec:introduction}

\begin{figure}[b]
	\centering
	\includegraphics[width=.9\linewidth]{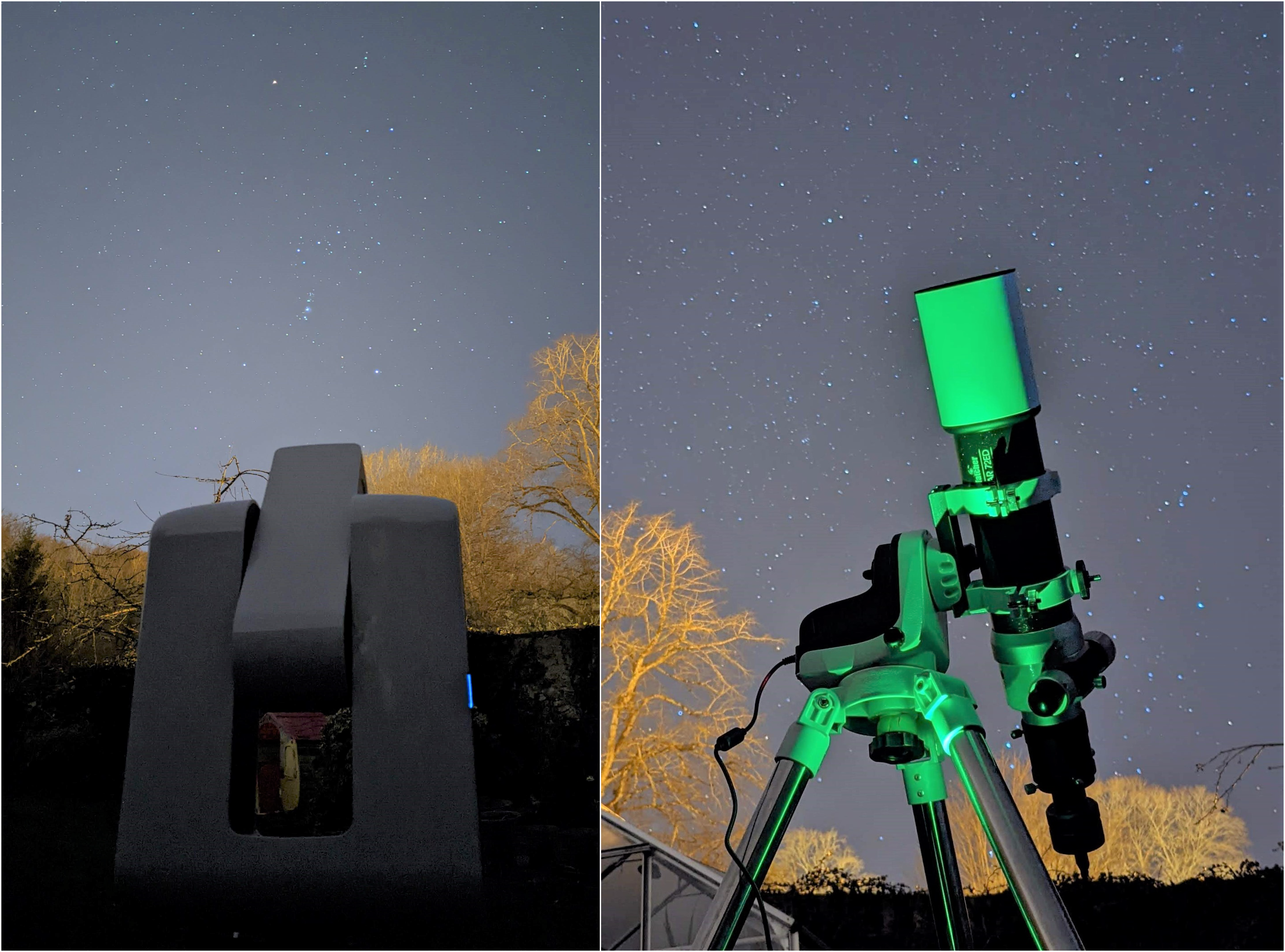}
	\caption{EAA setups used to capture data. The first one is a Stellina automated station, the second one is a 72/420 apochromatic refractor complemented by a low-end SVBONY SV305 digital camera -- connected to a laptop and driven by a dedicated software.} 
	\label{fig:setup}
\end{figure}

\noindent Nowadays, Electronically Assisted Astronomy (EAA) is widely applied by astronomers to observe deep sky objects (nebulae, galaxies, star clusters). 
By capturing raw images directly from a digital camera coupled to a telescope and applying lightweight image processing (fast alignment and stacking), this approach allows to generate enhanced views of deep sky targets that can be displayed in near real time on a screen (laptop, tablet, smartphone) (\figurename~\ref{fig:setup}). 

EAA also enables observing faint deep sky targets in difficult outdoor conditions, for example in geographical zones heavily impacted by light pollution or during a night with Moon (it considerably affects the sky background, often making it difficult the observations). 
Celestial objects like nebulae and galaxies are almost invisible through direct observation in an urban or suburban night sky; with EAA they become impressive and detailed \cite{eaa2022}.
In practice, hundreds of targets can be imaged -- they are listed in well-known astronomical catalogs (Messier, New General Catalog (NGC), Index catalog (IC), Sharpless, Barnard) and described many books and software \cite{zack2018software}.

Thus, a large quantity of images are handled by astronomers during such EAA sessions: the targets are numerous, the observation conditions variable, which means that each image is different.
Quality may depend on many parameters \cite{redfern2020astrophotography}, among them:
\begin{itemize}
\item Instrument: aperture and focal ratio, optical quality, digital camera sensitivity and read noise, tracking mount precision.
\item Setup installation: balance and stability of tripod, focusing, collimation.
\item Seeing conditions: light pollution weather (clouds, fog, wind), moon phase, steadiness and transparency of the atmosphere.
\end{itemize}
Some of these conditions may vary during the same night, meaning that acquired data may have very heterogeneous quality levels.

During night capture sessions, it is possible to visualize \emph{on live} the target by doing a lightweight processing: on-the-fly raw frames alignment and stacking, then quick cosmetic processing (in general, histogram stretching). 
This type of on-the-fly processing is performed by software such as SharpCap \footnote{\url{https://www.sharpcap.co.uk}.}
At the beginning, the stacked image is very noisy, then with the accumulation of raw images, this image will become more qualitative. 

After night capture sessions (generally days after), raw images are then heavily post-processed in dedicated editing / post-processing tools, allowing to mitigate most of the issues (noise, aberrations removal, blur) and to enhance the signal (contrast stretching, color correction) \cite{bracken2017deep,adake2022trust}.
The calibration images play an important role in this phase.
However, all these tools are very complex, and only very experienced users can really estimate the quality of the final processed images and thus improve it in a relevant way.

In this paper, we propose to combine Image Quality Assessment (IQA) and Automated Machine Learning (AutoML) to automatically rate astronomical RGB images: it aims at guiding EAA sessions and then images post-processing.

The rest of this article is organized as follows. 
Firstly, related works are described (Section \ref{sec:related}). 
Then, an approach with dataset preparation, model training and a prototype are detailed (Section \ref{sec:approach}). 
Finally, preliminary results on astronomical images are presented (Section \ref{sec:experiments}) and discussed (Section \ref{sec:discussion}). 
We conclude by opening some perspectives (Section \ref{sec:conclusion}).

\section{\uppercase{Related works}}
\label{sec:related}

\subsection{\uppercase{IQA}}

\noindent Quality of astronomical images is traditionally estimated through two methods:
\begin{itemize}
\item Signal-to-noise ratio (SNR): ratio of the strength of the astronomical signal to the level of the noise in an image. A higher SNR indicates that the image is of higher quality.
\item Full Width at Half Maximum (FWHM): sharpness of the point sources in an image, such as stars. A smaller FWHM indicates that the image is of higher quality.
\end{itemize}

An other popular measure is the highest magnitude of the faintest star/object visible in the image: it needs precise astrometry to do the comparison between the image and the known deep sky objects and stars present in celestial catalogues \cite{hogg2008automated}.

Recently, numerous IQA generic approaches were developed \cite{zhai2020perceptual}. 
In this paper, we focus on No-reference (NR) and Blind methods to rate single RGB images; among them we can list:
\begin{itemize}
	\item BRISQUE, efficient on natural scenes: a score between 0 (good quality) and 100 (poor quality) is computed \cite{mittal2012no}. It works well to filter really bad images by using a fixed threshold (i.e. 70).
	\item Deep Learning methods like NIMA (Neural Image Assessment): technical and aesthetic scores between 0 (bad) and 10 (good) \cite{talebi2018nima}. In practice, this score is not efficient on low-light images \cite{parisot2022applying}.
	\item Deep CNN-Based Blind Image Quality Predictor (DIQA) methodology: two models provide scores between 0 (bad) and 10 (good) \cite{jongyoo2019}.
\end{itemize}

Naively, these generic methods may be used to filter very bad astronomical images by using a value threshold, but it is not efficient in practice. 
Let's take the example of two different images of the M17 nebulae \cite{PARISOT2023109133} analyzed with two Python tools ( \emph{image-quality} package \footnote{\url{https://pypi.org/project/image-quality/}} and NIMA tensorflow model \footnote{\url{https://tinyurl.com/idealo-iqa}}).
In this typical case, the BRISQUE and NIMA evaluations are slightly better for the first image. 
However, the second image has a better overall quality, especially regarding contrast, luminance and noise.

\begin{figure}[h]
	\centering
	\includegraphics[width=.9\linewidth]{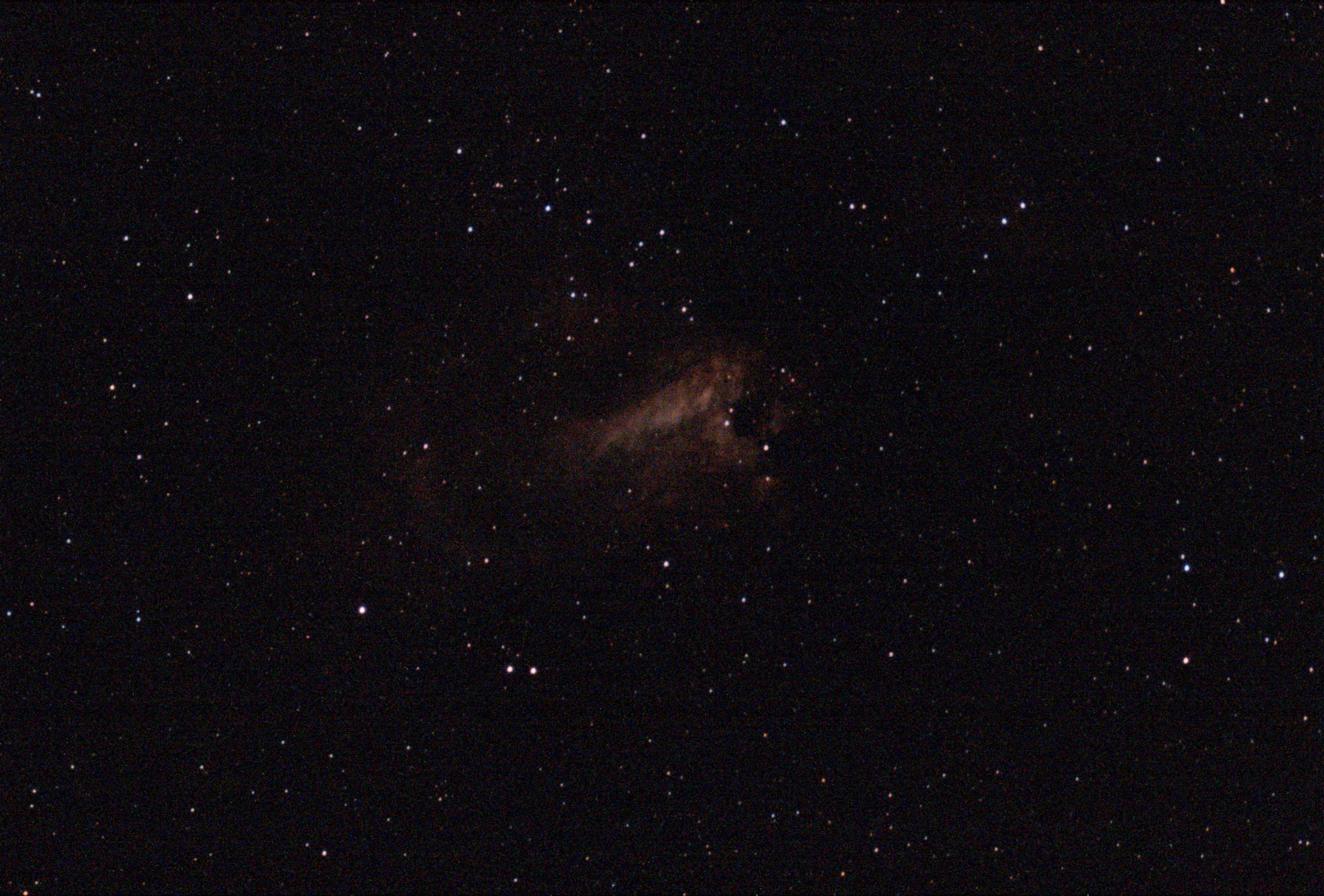}
	\caption{First image of M17 (aka Omega Nebula) - BRISQUE score is 28.25 and NIMA score is 4.52. The image lacks contrast, the stars are mixed with noise and the nebulosity is not visible.} 
	\label{fig:compare_iqa_1}
\end{figure}

\begin{figure}[h]
	\centering
	\includegraphics[width=.9\linewidth]{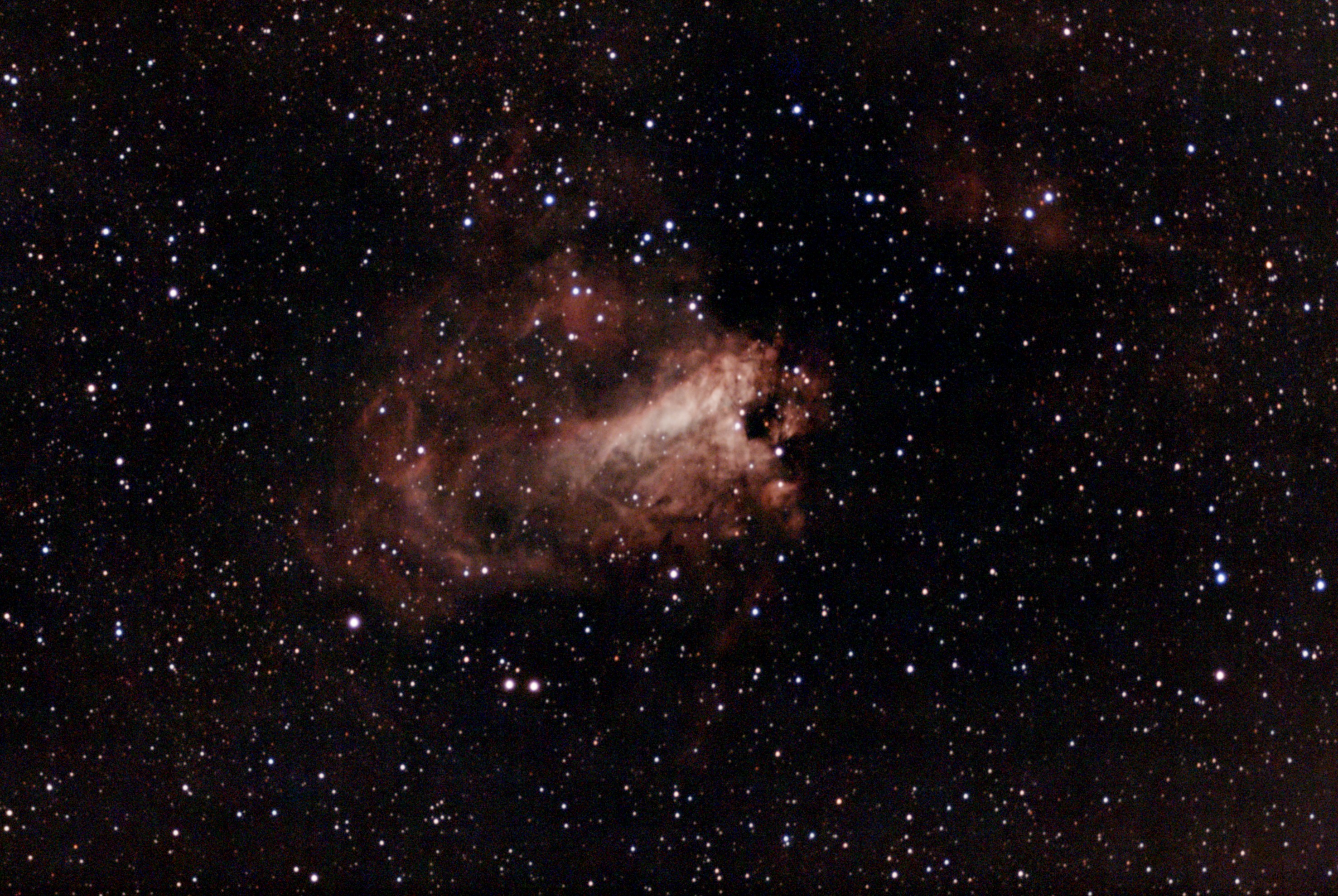}
	\caption{Second image of M17 - BRISQUE score is 28.28 and NIMA score is 4.17. The contrast is good, stars and nebulosity are clearly distinct from noise.} 
	\label{fig:compare_iqa_2}
\end{figure}

We can note that a recent IQA method based on clustering was proposed to deal with ground-based astronomical images captured by professional surveys \cite{teimoorinia2020assessment}.

\subsection{\uppercase{AutoML}}

AutoML consists in generating and deploying Machine Learning models from an input dataset with little or no configuration and coding effort \cite{hutter2019automated}.
The growing application of Machine Learning in a wide range of fields has led to the design of frameworks facilitating the production of readily actionable models.
Let us consider the traditional pipeline for traditional Machine Learning tasks:
\begin{itemize}
	\item Data preprocessing is required to adjust raw data to the specificity of Machine Learning algorithms \cite{zelaya2019towards}: cleansing, feature selection, sampling, transformation, etc.
	\item A Machine Learning model architecture is selected and then trained by using the prepared data \cite{raschka2018model}. 
	\item Depending on the algorithm, some hyper-parameters have to be optimized to improve the accuracy of the model; in general, this is realized through heuristics requiring heavy computation \cite{feurer2019hyperparameter}.
	\item Model accuracy is evaluated by computing standard statistical tests (AUC, Precision, Recall, F1, etc.) with a given strategy (holdout or cross-validation).
\end{itemize}
In practice, all those steps are time-consuming and exposed to methodological errors. 
AutoML platforms aim at systematizing the whole process in order to launch it a number of times with various combinations: numerous pipelines are tested, the obtained models are evaluated and the \emph{most accurate} one is finally selected \cite{raschka2018model}.


\section{\uppercase{Approach}}
\label{sec:approach}

Our approach consists in providing an \textit{image regressor} model to rate the quality of RGB astronomical images obtained during EAA sessions.
The model aims at taking into account the following criteria: contrast/luminance, noise and sharpness.
To this end, we have designed a set of images associated to a defined rating, and the task consists in training a model to fit with this score definition.
AutoML allows to test automatically a multitude of combinations -- leading to models with various characteristics: sizes (i.e. parameters count and feature map size, etc.) and accuracies.
Then, these models are tested through a process based on specific data augmentation process, in order to select the most robust model (\textit{test-time augmentation} \cite{shorten2019survey}).

\subsection{\uppercase{Data preparation}}

\noindent We have built a dataset with deep sky images and an associated rate (between 0 and 10, from \textit{bad} to \textit{good}), in a similar way to what is done in the DIQA methodology \cite{jongyoo2019}.

As original sources, we have used:
\begin{itemize}
\item Galaxy10 DECals Dataset containing 17736 256x256 pixels RGB galaxy images \cite{Leung_2018}. 
\item Nebula Dataset containing 1657 high-resolution images extracted from Wikimedia commons \cite{ravi2020}.
\end{itemize}

Then, we have prepared a set of ideal images, obtained after a long manual treatment of initial images with different editing software that are efficient to improve astronomical images: Siril, TopazLabs \cite{redfern2020astrophotography}. 
For each ideal image, a rating of 10 was assigned.
We have produced a set of transformed images by applying random degradations to modify noise level (adding Gaussian \& Poisson noise), sharpness (blurring, deforming stars) and luminance/contrast (adding background level and gradient, reducing signal, degrading color saturation).

Each transformed image is then rated using a value between 0 (bad quality) an 10 (good quality) . 
To determine a value, we have evaluated the impacts of distortions on contrast/luminance, noise and sharpness -- by comparing the transformed image with the ideal image.

\begin{itemize}
\item Contrast/luminance: we used Structural Similarity Index Measure (SSIM) because it is efficient to compare difference of contrast and luminance \cite{aliakhmet2019temporal}: 1 is given to a perfectly similar image and 0 indicates no similarity. 


\item Noise: we used the normalized Noise Variance difference between the ideal and the transformed image. Noise Variance is estimated through the Fast Estimation method.


\item Sharpness: we used the normalized FWHM difference between the ideal and the transformed image. FWHM is estimated through an heuristic based on stars detection \footnote{\url{https://tinyurl.com/starsfinder}}.
 

\end{itemize}

We have defined the final rating associated to the transformed image as:

$10*(r_{con} - max(r_{noi},A) - max(r_{sha},B)$

This formula ensures that each starting criterion is taken into account in the rating -- the interest being to have an index to compare the images as a whole, whatever the defect. 
The value $A$ and $B$ acts as maximum \emph{malus} associated to noise and lack of sharpness -- in practice we have empirically used 4 as value for this two constants.

\begin{figure}[h]
	\centering
	\includegraphics[width=.55\linewidth]{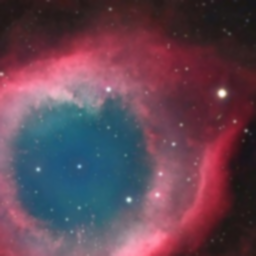}
	\caption{A good-quality 256x256 RGB image of Helix Nebula (NGC7293) -- rating: 9. The contrast is good, the noise level is low.} 
	\label{fig:data_good}
\end{figure}

\begin{figure}[h]
	\centering
	\includegraphics[width=.55\linewidth]{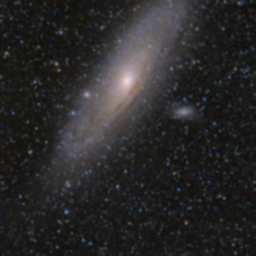}
	\caption{A 256x256 RGB image with moderated quality of the Andromeda galaxy (M31) -- rating: 6. Slight noise and blur have been added to degrade the original image.} 
	\label{fig:data_normal}
\end{figure}

\begin{figure}[h]
	\centering
	\includegraphics[width=.55\linewidth]{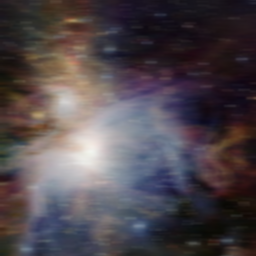}
	\caption{A poor-quality 256x256 RGB image of Orion Nebula (M42) in the built dataset -- rating: 2. Background noise and strong motion blur have been added to degrade the original image.} 
	\label{fig:data_bad}
\end{figure}

Thus, the constructed dataset contains several thousands of images of different qualities and associated ratings, for instance: good (Figure \ref{fig:data_good}), medium-quality (Figure \ref{fig:data_normal}), very bad (Figure \ref{fig:data_bad}).

The image resolution (256x256) is a good compromise because it corresponds to a standard amount of data for recent Deep Learning model architectures.


In the next section, we show how this dataset is then used to train a IQA model able to rate astronomical RGB images.

\subsection{\uppercase{Training}}
\label{sec:implementation}

\noindent To run AutoML, a Python prototype was implemented.
Image processing is realized with well-known open-source packages like \emph{openCV} \footnote{\url{https://pypi.org/project/opencv-python/}} and \emph{scikit-images} \footnote{\url{https://pypi.org/project/scikit-image/}}.

We have used AutoKeras -- an open source Python package based on Bayesian optimization \cite{JMLR:v24:20-1355}.
AutoKeras aims at building and fine-tuning Machine Learning and Deep Learning models by only defining inputs and expected outputs.
Other AutoML solutions exist (like TPOT \cite{olson2016tpot}), but AutoKeras provides a native support for \textit{Image Regression} models.

Behind the scenes, AutoKeras generates and launches numerous predefined training pipelines on-the-fly (model architecture selection, preprocessing, hyper-parameters setting, training, model evaluation and comparison).
it will notably check if normalization step of the input data will allow to obtain a better model. 
It will also test a whole list of hyper-parameters such as drop-our rate, activation function (ex: Relu, Sigmoid), optimization algorithm (ex: ADAM, RMSprop), learning rate, etc. 
AutoKeras does not work randomly to find the best configuration: it tests a number of pre-defined pipelines, then refines the best configuration by making small mutations (as an evolutionary algorithm would do) \cite{JMLR:v24:20-1355}. 
Step by step, the pipeline producing the best model is thus refined up to a user-defined limit (i.e. number of trials).

The computations were executed on a computing infrastructure with the following hardware capabilities: 40 cores and 128 GB RAM (Intel(R) Xeon(R) Silver 4210 CPU @ 2.20GHz) and NVIDIA Tesla V100-PCIE-32GB.
CUDA \footnote{\url{https://developer.nvidia.com/cuda-zone}} and NUMBA \footnote{\url{http://numba.pydata.org}} frameworks have been used in background to optimize the hardware usage during images treatment(CPUs and GPUs).

\begin{figure}[h]
	\centering
	\includegraphics[width=.9\linewidth]{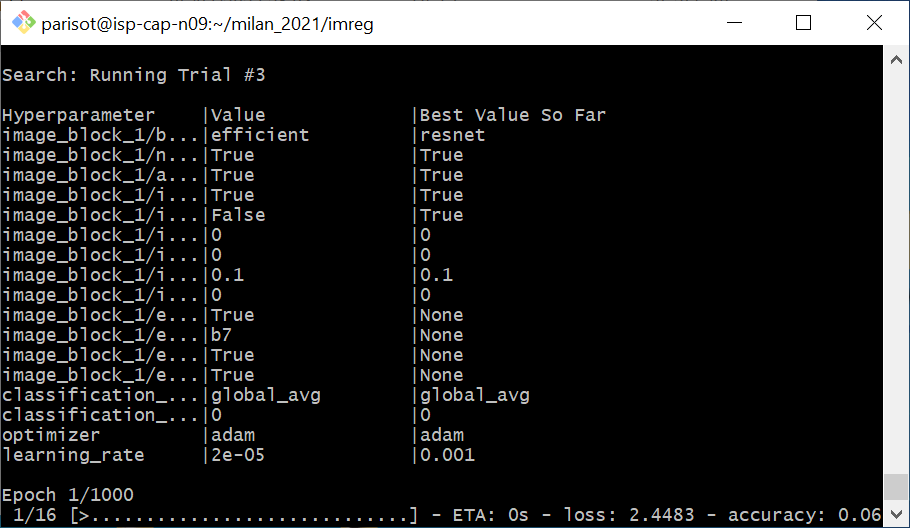}
	\caption{Console dashboard of AutoKeras during the executions of numerous training pipelines. It is thus possible to monitor the execution of the pipelines and to see which is the best model at time \emph{t}.} 
	\label{fig:autokeras_dashboard}
\end{figure}

After numerous experiments, we have run thousands of different pipelines by combining variations of data preprocessing, optimizers usage, different hyper-parameters values (\figurename~\ref{fig:autokeras_dashboard}).

Two models were shown to be worthy of interest after all the calculations:
\begin{itemize}
\item The best one leds to a model embedding a \emph{ResNet50} model \cite{he2016deep}, with 23 millions trainable parameters and 53 120 non-trainable parameters -- obtained with a ADAM optimizer (learning rate of 0.01). Its accuracy is 1.19 (Mean Squared Error, MSE) on the test dataset. The R squared value is 0.4 -- which is rather satisfying because we have here a regression on images.
\item The smaller model is based on \emph{EfficientNetB1} with 6.589.337 parameters \cite{tan2019efficientnet} -- obtained with a ADAM optimizer (learning rate of 0.00001). The model accuracy was worse (MSE 1.25).
\end{itemize}

To observe the robustness of these two trained models on realistic data, we tested them on a augmented test dataset -- on images having additional distortions -- not present in the training set.


\subsection{\uppercase{Model selection on an augmented test set}}

To obtain an additional large and realistic test dataset, we wrote several Python scripts for realistic image augmentation, i.e. to reproduce defects that are frequently found in astronomical images. 
For example, we have added different types of noises (Gaussian, Poisson, Salt and Pepper), we have merged the real signal with a realistic sky background \cite{bradley2016photutils}.
We also have blurred images in a sophisticated way by using image augmentation techniques based on Deep Learning -- especially for motion blur \cite{jung2019imgaug}.
In a similar way, we generated starless versions of the images, in order to test the robustness of the model -- this was realized with a dedicated Deep Learning model \footnote{\url{https://www.starnetastro.com}}.


Both models are globally able to reproduce the defined rating -- to make an estimation between images with and without defects.
However the IQA model based on ResNet50 provides much better results, especially on the ratings of very good or very bad images (around 0 or around 10).

\begin{figure}[h]
	\centering
	\includegraphics[width=.9\linewidth]{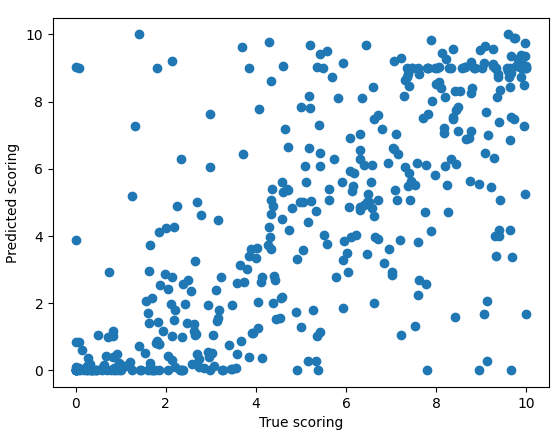}
	\caption{Regression plot of expected (x axis) and computed ResNet50 IQA ratings (y axis) obtained on a part of the augmented dataset (i.e. 100 images).} 
	\label{fig:regression_plot}
\end{figure}

\begin{table}
	\centering
	\caption{Rating obtained with the best ResNet50 IQA model on a augmented test dataset built from 500 different 256x256 RGB images. The (mean, standard deviation, minimum, maximum) outputs of the model are given for each images set. }	
	\label{table:results0}
	\begin{tabular}{|l|r|r|r|r|}
		\hline
		            		   & \multicolumn{4}{c|}{ResNet50 ratings}   \\
		\hline
		            		   & mean  & std  & min  & max   \\
		\hline
		No distortion  	   & 9.2   & 0.6  & 3.6  & 10    \\
		\hline
		Distortion  	   	   & 3.6   & 0.9  & 1.1  & 6.9     \\
		\hline
		Strong distortion  & 0.7   & 0.6  & 0    & 3.6  \\
		\hline             
	\end{tabular}
\end{table}

\begin{table}
	\centering
	\caption{Rating obtained with the best EfficientNetB1 IQA model on a augmented test dataset built from 500 different 256x256 RGB images. The (mean, standard deviation, minimum, maximum) outputs of the IQA model are given for each images set. }	
	\label{table:results1}
	\begin{tabular}{|l|r|r|r|r|}
		\hline
		            		   & \multicolumn{4}{c|}{EfficientNetB1 ratings}   \\
		\hline
		            		   & mean  & std  & min  & max   \\
		\hline
		No distortion  	   & 8.9   & 1.8  & 1.2  & 10    \\
		\hline
		Distortion  	   	   & 5.5   & 2.8  & 0.7  & 10      \\
		\hline
		Strong distortion  & 2.9   & 3.2  & 0    & 10  \\
		\hline             
	\end{tabular}
\end{table}

Certainly, there are more powerful classification architectures than ResNet50 and EfficientNet. 
Nevertheless, AutoML allowed us to obtain optimized models for the use-case presented in this paper.

In the next section, we detail the efficiency of the ResNet50 IQA model on live stacked images captured during EAA sessions.

\section{\uppercase{Experiments}}
\label{sec:experiments}

The model was then tested on data captured during EAA sessions with two setups:
\begin{itemize}
\item 300 live stacked images obtained from 10 seconds sub-frames, with short total integration time (from 20 to 30 minutes) by using a Stellina observation station \cite{PARISOT2023109133}. Images correspond to different deep sky objects (example: Messier 31, NGC4565, etc.).
\item 100 live stacked images obtained from 5 seconds sub-frames, with short total integration time (from 20 to 30 minutes) by using a 72/420 refractor and a dedicated low-end astronomical digital camera(\footnote{\url{https://tinyurl.com/sv305}}).
\end{itemize}

All these RGB images have a high resolution (3096x2048 for the first set, 2048x2048 for the second set). 
As the input of our IQA model are 256x256 RGB images, all of them were split into patches and the overall IQA rating is the mean of patch ratings (no overlap).

\begin{table}
	\centering
	\caption{Evaluation of live stacked images captured during EAA sessions with two setups: a Stellina observation station (300 images) and a 72/400 refractor coupled to a low-end digital camera (100 images). The (mean, standard deviation, minimum, maximum) outputs of the ResNet50 IQA model are listed.}	
	\label{table:results2}
	\begin{tabular}{|l|r|r|r|r|}
		\hline
		            		   & \multicolumn{4}{c|}{ResNet50 IQA rating}   \\

		\hline
							   & mean  & std  & min  & max   \\
		\hline
		Stellina  		   	   & 5.8  & 1     & 2.6 & 9       \\
		\hline
		72/400 refractor  	   & 1.3  & 0.7   & 0.6 & 3.1     \\
		\hline             
	\end{tabular}
\end{table}

The computed ratings are representative of the image sets: in practice, the Stellina observation station provides much better quality images than the other setup (Table \ref{table:results2}).
They are useful to compare images too: the second M17 nebula image presented in Figure \ref{fig:compare_iqa_1} has a rating which is 50\% higher than the rating of the first image (Figure \ref{fig:compare_iqa_2}).

\begin{figure}[h]
	\centering
	\includegraphics[width=.8\linewidth]{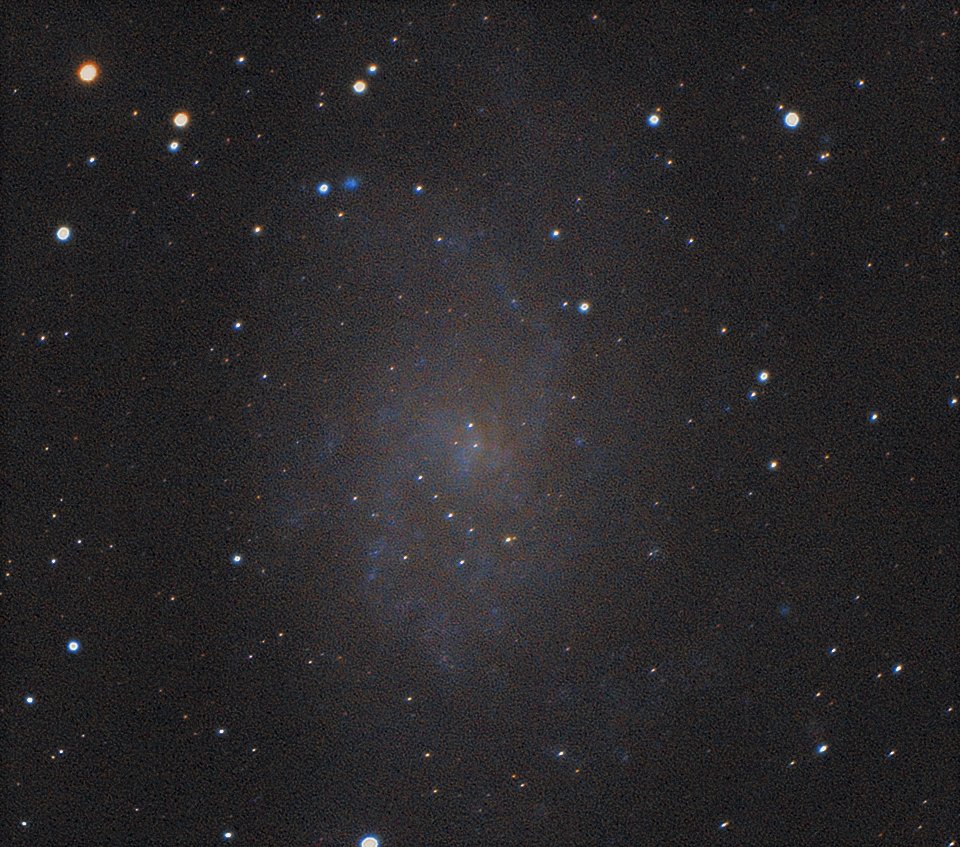}
	\caption{Live stacked image of the M33 galaxy obtained with a SV305 camera and a 72/400 refractor: high noise, malformed stars and insufficient contrast. The ResNet50 IQA model rating is 1.49.} 
	\label{fig:res_iqa_1}
\end{figure}

\begin{figure}[h]
	\centering
	\includegraphics[width=.8\linewidth]{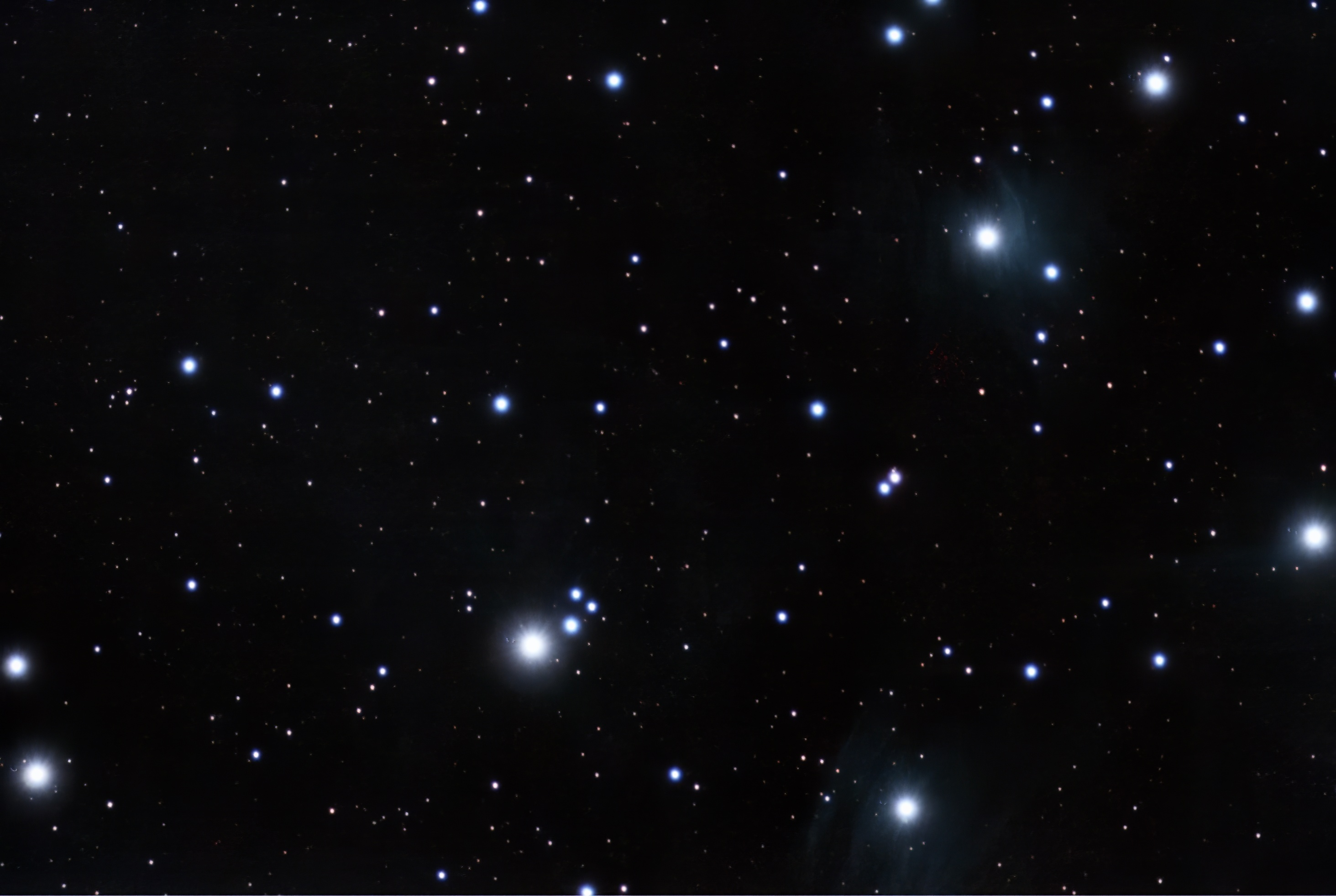}
	\caption{Live stacked image of Pleiades (M45) captured with a Stellina automated station: with little noise and stars with a punctual appearance. The ResNet50 IQA model rating is 7.1 .} 
	\label{fig:res_iqa_2}
\end{figure}

\begin{figure}[h]
	\centering
	\includegraphics[width=.9\linewidth]{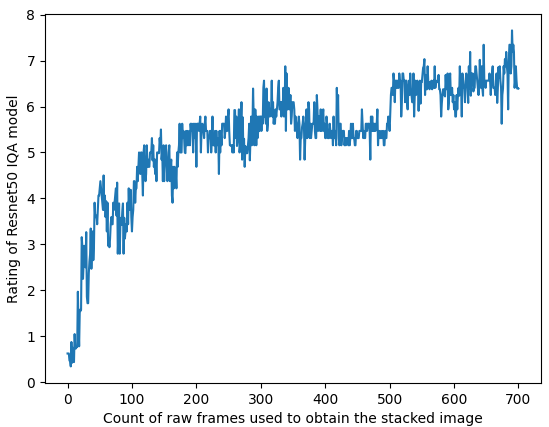}
	\caption{Evolution of IQA score during a EAA session for observing the California Nebula (NGC1499). The more raw images accumulated, the better the final stacked image quality.} 
	\label{fig:res_live_ngc1499}
\end{figure}

One of the interests of the approach is to measure the quality of the stacked image obtained during a EAA capture session. 
In theory, an astronomical image improves when accumulating integration time (i.e. by collecting as much data as possible).
Capturing data is time-consuming and sometimes challenging (especially due to weather conditions), and the IQA estimation on the live stacked image may help to capture only what is necessary to have a stacked image with the desired quality (Figure \ref{fig:res_live_ngc1499}).
As the number of raw images increases, it becomes more difficult to increase the quality of the stacked image.
Let's take the example of the observation of NGC1499 with a Stellina station: 711 images of 10 seconds of exposure time each \footnote{\url{https://youtu.be/BTURaF9dTIU}}. 
The graph shows that the quality increases strongly at the beginning of the capture, and less afterwards (Figure \ref{fig:res_live_ngc1499}). 
The score can thus be used to determine if it is still relevant to continue data acquisition.

Out of curiosity, we have tested the IQA model on two extreme cases of well-known astronomical pictures.
Although not typical use cases of our approach, it gives us an overview of what our model produces on different images:
\begin{itemize}
\item First Andromeda image captured by Isaac Roberts in (1899) \footnote{\url{https://commons.wikimedia.org/wiki/File:Pic_iroberts1.jpg}}: our IQA model provides a rating of 2.56. Even if the image is incredible for its time, it is far from the standards expected today: a poor score is logic.
\item Webb's First Deep Field -- a long-exposure image of the SMACS0723 galaxy cluster captured vy the James Webb Space Telescope
 \footnote{\url{https://tinyurl.com/webbdf}}: our IQA model computes a rating of 7.2 for the part without the JWST's typical diffraction spikes \cite{rigby2022characterization}. The rating drops to 4.9 with these spikes because the model seems to consider them as defects.
\end{itemize}

\section{\uppercase{Discussion}}
\label{sec:discussion}

With our approach, we can observe that the image defects are really penalized by the IQA model as we have chosen to mix potential distortions into the training process. 
There is a drawback: this method is not able to list, grade and locate precisely the defects that are most present in an image, and for this we will have to go further (for instance, by combining with object detection).

Another point concerns the execution time of the IQA model on large images. 
Even if training is realized on a high performance computing platform, the models use should be possible on normal computers with modest capabilities -- especially for on-the-fly evaluation of stacked images.

Let's take the example of a 4096x4096 astronomical image: with no overlap, we may need to evaluate 256 256x256 patches -- it may take some time depending of the hardware.
To be efficient, one must try to minimize the number of calculations required. 
In a pragmatic way, the following strategies can be applied:
\begin{itemize}
\item Decrease the resolution of the image to reduce the count of patches to evaluate.
\item Estimate the IQA rating of a small relevant subset of patches -- for instance by ignoring dark patches or patches with low signal.
\end{itemize}
During our experiments, we have concluded that the second one provides better results.
Further performance optimizations will be realized after deep analysis of model execution with dedicated tools \cite{9150136}.

\section{\uppercase{Conclusion}}
\label{sec:conclusion}

\noindent This paper presented an approach to automatically estimate the quality of astronomical RGB images.
A dedicated model was trained by using Automated Machine Learning, and then tested on various image sets captured during Electronically Assisted Astronomy sessions.
A Python prototype was presented and preliminary results were discussed.
In future works, we will extend the approach building on these current results to design and train additional models that are both more sophisticated and interpretable, by relying on eXplainable Artificial Intelligence.
\\

\FloatBarrier

\noindent \textbf{Acknowledgments}: 
This research was funded by the Luxembourg National Research Fund (FNR), grant reference 15872557. 
Tests were realized on the LIST AIDA platform, thanks to Raynald Jadoul and Jean-François Merche.

\bibliographystyle{apalike}
{\small
\bibliography{article}}

\begin{thebibliography}{}

\bibitem[Adake, 2022]{adake2022trust}
Adake, N. (2022).
\newblock Trust the process: An investigation into astrophotography.
\newblock In {\em AIAA Southeastern Regional Student Conference}.

\bibitem[Aliakhmet and James, 2019]{aliakhmet2019temporal}
Aliakhmet, K. and James, A.~P. (2019).
\newblock Temporal g-neighbor filtering for analog domain noise reduction in
  astronomical videos.
\newblock {\em IEEE Trans. on Circuits and Systems II: Express Briefs},
  66(5):868--872.

\bibitem[Bracken, 2017]{bracken2017deep}
Bracken, C. (2017).
\newblock The deep-sky imaging primer.
\newblock {\em Deep-sky Publishing}.

\bibitem[Bradley et~al., 2016]{bradley2016photutils}
Bradley, L., Sipocz, B., Robitaille, T., Tollerud, E., Deil, C., Vin{\'\i}cius,
  Z., Barbary, K., G{\"u}nther, H.~M., Bostroem, A., Droettboom, M., et~al.
  (2016).
\newblock Photutils: Photometry tools.
\newblock {\em Astrophysics Source Code Library}, pages ascl--1609.

\bibitem[Feurer and Hutter, 2019]{feurer2019hyperparameter}
Feurer, M. and Hutter, F. (2019).
\newblock Hyperparameter optimization.
\newblock In {\em Automated Machine Learning}, pages 3--33. Springer, Cham.

\bibitem[He et~al., 2016]{he2016deep}
He, K., Zhang, X., Ren, S., and Sun, J. (2016).
\newblock Deep residual learning for image recognition.
\newblock In {\em IEEE CVPR 2016}, pages 770--778.

\bibitem[Hogg et~al., 2008]{hogg2008automated}
Hogg, D.~W., Blanton, M., Lang, D., Mierle, K., and Roweis, S. (2008).
\newblock Automated astrometry.
\newblock In {\em Astronomical Data Analysis Software and Systems XVII}, volume
  394, page~27.

\bibitem[Hutter et~al., 2019]{hutter2019automated}
Hutter, F., Kotthoff, L., and Vanschoren, J. (2019).
\newblock {\em Automated machine learning: methods, systems, challenges}.
\newblock Springer Nature.

\bibitem[Jin et~al., 2023]{JMLR:v24:20-1355}
Jin, H., Chollet, F., Song, Q., and Hu, X. (2023).
\newblock Autokeras: An automl library for deep learning.
\newblock {\em Journal of Machine Learning Research}, 24(6):1--6.

\bibitem[Jin and Finkel, 2020]{9150136}
Jin, Z. and Finkel, H. (2020).
\newblock Analyzing deep learning model inferences for image classification
  using {OpenVINO}.
\newblock In {\em IPDPSW 2020}, pages 908--911.

\bibitem[Jung, 2019]{jung2019imgaug}
Jung, A. (2019).
\newblock Imgaug documentation.
\newblock {\em Readthedocs. io, Jun}, 25.

\bibitem[Kim et~al., 2019]{jongyoo2019}
Kim, J., Nguyen, A.-D., and Lee, S. (2019).
\newblock Deep cnn-based blind image quality predictor.
\newblock {\em IEEE Trans. on Neural Networks and Learning Systems},
  30(1):11--24.

\bibitem[Leung and Bovy, 2018]{Leung_2018}
Leung, H.~W. and Bovy, J. (2018).
\newblock Deep learning of multi-element abundances from high-resolution
  spectroscopic data.
\newblock {\em Monthly Notices of the Royal Astronomical Society}.

\bibitem[Mittal et~al., 2012]{mittal2012no}
Mittal, A., Moorthy, A.~K., and Bovik, A.~C. (2012).
\newblock No-reference image quality assessment in the spatial domain.
\newblock {\em IEEE Trans. on image processing}, 21(12):4695--4708.

\bibitem[Olson and Moore, 2016]{olson2016tpot}
Olson, R.~S. and Moore, J.~H. (2016).
\newblock {TPOT: A tree-based pipeline optimization tool for automating machine
  learning}.
\newblock In {\em Workshop on automatic machine learning}, pages 66--74. PMLR.

\bibitem[Parisot et~al., 2022]{eaa2022}
Parisot, O., Bruneau, P., Hitzelberger, P., Krebs, G., and Destruel, C. (2022).
\newblock Improving accessibility for deep sky observation.
\newblock {\em {ERCIM} News}, 2022(130).

\bibitem[Parisot et~al., 2023]{PARISOT2023109133}
Parisot, O., Hitzelberger, P., Bruneau, P., Krebs, G., Destruel, C., and
  Vandame, B. (2023).
\newblock {MILAN Sky Survey, a dataset of raw deep sky images captured during
  one year with a Stellina automated telescope}.
\newblock {\em Data in Brief}, 48:109133.

\bibitem[Parisot and Tamisier, 2022]{parisot2022applying}
Parisot, O. and Tamisier, T. (2022).
\newblock Applying genetic algorithm and image quality assessment for
  reproducible processing of low-light images.
\newblock In {\em IMPROVE 2022}, pages 189--194.

\bibitem[Raschka, 2018]{raschka2018model}
Raschka, S. (2018).
\newblock Model evaluation, model selection, and algorithm selection in machine
  learning.

\bibitem[Ravi, 2020]{ravi2020}
Ravi, A. (2020).
\newblock Nebula images.

\bibitem[Redfern, 2020]{redfern2020astrophotography}
Redfern, G.~I. (2020).
\newblock {\em Astrophotography is Easy!: Basics for Beginners}.
\newblock Springer.

\bibitem[Rigby et~al., 2022]{rigby2022characterization}
Rigby, J., Perrin, M., McElwain, M., Kimble, R., Friedman, S., Lallo, M.,
  Doyon, R., Feinberg, L., Ferruit, P., Glasse, A., et~al. (2022).
\newblock Characterization of jwst science performance from commissioning.
\newblock {\em arXiv preprint arXiv:2207.05632}.

\bibitem[Shorten and Khoshgoftaar, 2019]{shorten2019survey}
Shorten, C. and Khoshgoftaar, T.~M. (2019).
\newblock A survey on image data augmentation for deep learning.
\newblock {\em Journal of big data}, 6(1):1--48.

\bibitem[Talebi and Milanfar, 2018]{talebi2018nima}
Talebi, H. and Milanfar, P. (2018).
\newblock Nima: Neural image assessment.
\newblock {\em IEEE Trans. on Image Processing}, 27(8):3998--4011.

\bibitem[Tan and Le, 2019]{tan2019efficientnet}
Tan, M. and Le, Q. (2019).
\newblock Efficientnet: Rethinking model scaling for convolutional neural
  networks.
\newblock In {\em ICML}, pages 6105--6114. PMLR.

\bibitem[Teimoorinia et~al., 2020]{teimoorinia2020assessment}
Teimoorinia, H., Kavelaars, J., Gwyn, S., Durand, D., Rolston, K., and
  Ouellette, A. (2020).
\newblock Assessment of astronomical images using combined machine-learning
  models.
\newblock {\em The Astronomical Journal}, 159(4):170.

\bibitem[Zack et~al., 2018]{zack2018software}
Zack, M., Gannon, A., and McRoberts, J. (2018).
\newblock Software and apps to help the suburban astronomer.
\newblock In {\em Stargazing Under Suburban Skies}, pages 341--350. Springer.

\bibitem[Zelaya, 2019]{zelaya2019towards}
Zelaya, C. V.~G. (2019).
\newblock Towards explaining the effects of data preprocessing on machine
  learning.
\newblock In {\em ICDE 2019}, pages 2086--2090. IEEE.

\bibitem[Zhai and Min, 2020]{zhai2020perceptual}
Zhai, G. and Min, X. (2020).
\newblock Perceptual image quality assessment: a survey.
\newblock {\em Science China Information Sciences}, 63(11):1--52.

\end{thebibliography}

\end{document}